\documentclass[a4paper,12pt]{article}

\usepackage{graphicx} % Required for inserting images
\usepackage[cp1251]{inputenc} %кодировка
\usepackage[T1]{fontenc}   % Латинские и кириллические символы
\usepackage[english]{babel} %для нормального шрифта
\usepackage{indentfirst} %Красная строка первого абзаца
\usepackage{amsmath} %чтобы работала нумерация
\usepackage{float} %чтобы работала нумерация
\usepackage{caption}  %заголовки плавающих объектов

\numberwithin{equation}{section} %Нормальная нумерация

\title{High-energy behaviour of Fermi theory}
\author{A.T.Borlakov and D.I.Kazakov}

\date{}

\begin{document}

\maketitle
\begin{center}Bogoliubov Laboratory of Theoretical Physics\\ Joint Institute for Nuclear Research
\end{center}

\begin{abstract}
We consider the 4-fermion scattering amplitude in massless Fermi theory. Based on the Bogolyubov-Parasyuk theorem, which guarantees locality of the counter terms, we derive the recurrence relations for ultraviolet divergences of diagrams that establish a connection between successive orders of perturbation theory. We check their validity up to three loops comparing them with explicit calculation made earlier. Then we construct the corresponding RG equation that sums up the leading logarithmic contributions in all orders of perturbation theory. Numerical analysis of these equations in the asymptotic regime $s\sim t\sim u \sim E^2 \to \infty$ is performed for two cases: the unit and the V-A operator in the fermion current. We found out that for the unit operator the high energy behaviour of the theory in the leading order is characterized by the presence of the Landau pole, while for the V-A operator the theory is asymptotically free. Therefore, in the latter case, radiative corrections restores unitarity, which is violated at the tree level. We compare the obtained behaviour of the amplitude with one in the theory with the intermediate gauge bosons and found an overlap between them.
\end{abstract}

\section{Introduction}
It is well known that Fermi theory described by the Lagrangian
\begin{equation}
{\cal L}=i \bar \Psi \hat \partial \Psi - \frac{G}{\sqrt{2}} (\bar \Psi\hat{\cal O}_i \Psi) (\bar \Psi\hat{\cal O}_i \Psi),
\end{equation}
where the operator $\hat{\cal O}_i$ is an arbitrary combination of five operators, $ {\cal O}_i=1,\gamma^5,\gamma^\mu,\gamma^5\gamma ^\mu,\gamma^\mu\gamma ^\nu$, serves as an effective low-energy theory of weak interactions~\cite{GG}. From a physical point of view, the most interesting case is the $V-A$ interaction, when the operator $\hat{\cal O}=\gamma^\mu(1-\gamma^5)/2$. In the Standard Model at high energies, Fermi theory is replaced by the gauge theory of weak interactions with intermediate vector bosons. One of the reasons for this is the unitarity violation, which arises due to the increase in the amplitude of the 4-fermion interaction with energy already at the tree-level\cite{GG}. Another reason is that Fermi theory is non-renormalizable starting from with one loop. Ultraviolet divergences, which cannot be controlled using standard methods of renormalization theory and renormalization group, do not allow one to correctly determine the behaviour of the theory on a high-energy scale.

However, in a series of papers~\cite{BKTV,KBTV,KBBTV}, we showed how to work with non-renormalizable theories using the example of maximally supersymmetric gauge theories in higher dimensions. The corresponding renormalization group equations were obtained that allow summarizing the leading logarithms contributing to the scattering amplitude in the same way as is done in renormalizable theories.    

Here we apply the developed formalism to Fermi theory in four dimensions. This article is devoted to the summation of the leading UV divergences in all orders of perturbation theory for the amplitude of antifermion-fermion scattering based on the generalized renormalization group equations and the analysis of high-energy behaviour. The cases of unit $\hat{\mathcal{O}} = 1$ and vector-axial $\hat{\mathcal{O}}=\frac{1}{2}\gamma^\mu (1-\gamma^5)$ operators are considered. It is demonstrated that radiative corrections summed up over all orders of perturbation theory can significantly modify the behaviour of the amplitude and in some cases lead to its decrease with energy, this way restoring unitarity.

The article is organized as follows. Section~2 is devoted to the procedure of the R-operation, which is a method for eliminating ultraviolet divergences from multi-loop Feynman diagrams~\cite{BSh,Hepp,Zimmer}. Further, based on the Bogolyubov-Parasyuk theorem~\cite{BP,APZ}, which states that counter-terms are always local, we derive the recurrence relations for ultraviolet divergences of diagrams that establish a connection between successive orders of perturbation theory. In the third section, explicit formulas for the recurrence relations in Fermi theory are obtained. Their fulfillment with an accuracy up to the three-loop approximation is carried out by comparing with explicit calculations of Feynman diagrams. Then, based on the recurrence relations, the renormalization group equations for the scattering amplitudes are obtained. Section~4 presents an analysis of the asymptotic behaviour of  perturbation theory series for sufficiently high orders, as well as a numerical solution of the above-mentioned renormalization group equations in the high-energy domain. In summary, we present physical consequences and conclusions.

\section{The Bogolyubov-Parasyuk R-operation}
\begin{figure}[ht]
\begin{center}
\includegraphics[scale=0.52]{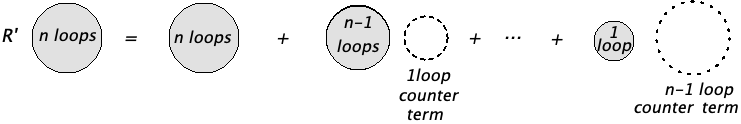}
\caption{Action of the $R'$-operation on an n-loop diagram \label{RO}}
\end{center}
\end{figure}

Our approach to summing the leading asymptotics of Feynman diagrams for the scattering amplitude
is based on the properties of the operation that eliminats ultraviolet divergences, the so-called $R$ operation, which is
valid in any local quantum field theory including non-renormalizable interactions. Define the $R$-operation acting on the diagram $G$ in an arbitrary order of perturbation theory as ~\cite{BSh}
\begin{equation}
R G= (1- K R')G,
\end{equation}
where $K$ is an operator that extracts the singular part, and $R'$ is an incomplete $R$ operation that eliminates divergences in the subgraphs of this diagram. For one-particle irreducible diagrams, it is defined as follows~\cite{Vasiliev, Vladimirov},
\begin{equation}
   R' = 1 - \sum_{\gamma}L\gamma + \sum_{\gamma, \gamma'}L\gamma\,L\gamma'-...;
   \label{R}
\end{equation}
where the operation $L\gamma=KR\gamma'$ replaces the subgraph $\gamma$ with its counter-term $L\gamma$. Schematically, the action of the $R'$ operation is shown in Figure~\ref{RO}.

\begin{figure}[ht]
\center{\includegraphics[scale=0.4]{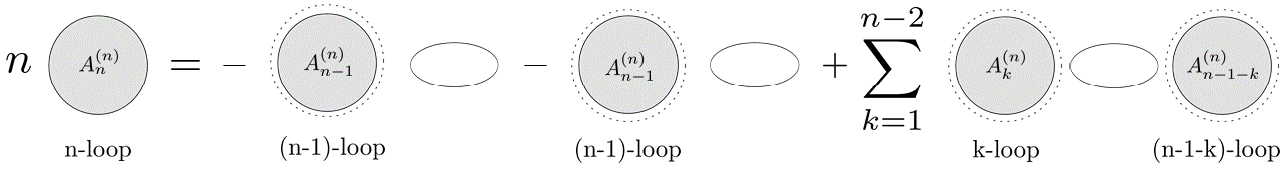}}
\caption{Recurrence relation for a four-point diagram \label{rec}}
\end{figure}

In what follows, to regularize the ultraviolet divergences, we use dimensional regularization, i.e. we calculate the integrals in dimension $4-2\epsilon$. In this case, the UV divergences have the form of poles in $\epsilon$, and in the n-loop diagram the highest pole is the pole of the nth order. If we limit ourselves only to the leading divergences, the action of the $R'$-operation on the amplitude in the nth order of perturbation theory results in  the following sequence~\cite{KBBTV}:
\begin{equation}
    \frac{A_n^{(n)}(\mu^2)^{n\epsilon}}{\epsilon^n} + \frac{A_{n-1}^{(n)}(\mu^2)^{(n-1)\epsilon}}{\epsilon^n}+...+ \frac{A_1^{(n)}(\mu^2)^{\epsilon}}{\epsilon^n}, 
    \label{An}
\end{equation}
where $A_n^{(n)}(\mu^2)^{n\epsilon}/\epsilon^n$ is the contribution of the diagram itself, and terms like $A_k^{(n)}(\mu^2)^{k\epsilon}/\epsilon^n$ follow from the $k$-loop subgraph after subtracting the $(n-k)$-loop counter-term. Expression (\ref{An}), according to the Bogolyubov-Parasyuk theorem, must be completely local, so it should not contain terms of the type $log^k(\mu^2)/\epsilon^m$ for all $k,m>0$, being decomposed into a series in $\epsilon$. This requirement yields $n-1$ equations for n coefficients $A_i^{(n)}$. Expressing the solutions in terms of the last coefficient $A_1^{(n)}$ corresponding to the one-loop diagram,  for the highest divergence of the n-loop diagram we obtain~\cite{KBBTV}:
\begin{equation}
    A_n^{(n)}=(-1)^{n+1}\frac{A_1^{(n)}}{n}\quad,
\end{equation}
and similarly for the total n-loop singularity:
\begin{equation}
    \sum_{k=1}^n A_k^{(n)}=\frac{A_1^{(n)}}{n}=(-1)^{n+1} A_n^{(n)}\quad.
\end{equation}
Thus, to find the leading divergence $\sim 1/\epsilon^n$, it is sufficient to know the divergence of the one-loop diagram multiplied by the $(n-1)$ counter-term. This circumstance makes it possible to write a recurrence relation connecting the divergences in subsequent orders of perturbation theory. 
For a four-point diagram, this recurrence relation is schematically shown in Figure~\ref{rec}. The first term here corresponds to the case when the remaining one-loop diagram is located at the edge, and the second one is when it is in the middle. The latter case leads to non-linearity in the recurrence relations and occurs starting from three loops.

The resulting recurrence relation should be interpreted correctly. The live Feynman diagram here shown by the bold line is only one-loop. The dotted line encircles the counter-term, which is either a constant (renormalizable theory) or a polynomial of external momenta. But the external lines of the $(n-1)$-loop diagram are partially internal to the $n$-loop one and, thus, the integration over these momenta is needed. Therefore, the resulting recurrence relation is algebraic in the renormalizable case but is integral in the non-renormalizable one. But in any case, the recurrence relation allows one to calculate the leading divergences (the highest pole in $\epsilon$), starting with the one loop without calculating the corresponding diagrams. The expicit form of these relations for the case of the four-fermion scattering will be presented in the next section.

\section{Recurrence relations for the four-fermion amplitude}

Our aim is to calculate the amplitude of the four-fermion scattering shown in Figure.~\ref{amp}: 

\begin{figure}[htb]
\begin{center}
\includegraphics[scale=0.6]{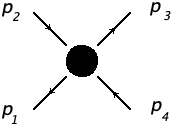}
\caption{Feynman diagram for four-fermion scattering. All momenta are incoming here.
}\label{amp}
\end{center}
\end{figure}

For this purpose, it is very convenient to use the spinor-helicity formalism ~\cite{Elvang, 1001, Schwinn}. The starting point for deriving recurrence relations is the leading ultraviolet divergences calculated in \cite{Borlakov} up to three loops. Let us start with the case of the unit operator. In the tree-level approximation, we have two structures, which are composed of Lorentz-invariant spinor products (in spinor-helicity notation):

\begin{equation*}
A_4^{(0)} = \langle 12 \rangle [34] - \langle 14 \rangle [32]\quad.
\label{A0}
\end{equation*}

As shown in \cite{Borlakov}, the scattering amplitudes proportional to these tree-level structures can be obtained from one another by simply replacing momenta $p_2\leftrightarrow p_4$ and  changing the sign due to the anticommutation. The recurrence relations reproducing the leading divergences in these amplitudes are independent. As a result, we will consider the amplitude proportional to the structure $\langle 12\rangle [34]$ without loss of generality. The contribution comes from the diagrams in the s-, t-, and u-channels, which are interconnected by crossing symmetry. For one, two, and three loops, the leading divergences are equal, respectively~\cite{Borlakov}(hereafter the total multiplier $\langle 12\rangle [34]$ is omitted):
\begin{eqnarray}
    S_1(s,t,u) &=&\frac{2s-t-u}{6\epsilon},\ T_1(s,t,u) =-\frac{t}{3\epsilon},\ U_1(s,t,u) =-\frac{u}{3\epsilon};  
    \label{R1stu}\\
    S_2(s,t,u)& =&\frac{5s^2}{12\epsilon^2}, \ T_1(s,t,u) =\frac{5t^2}{24\epsilon^2}, \ U_1(s,t,u) =\frac{5u^2}{24\epsilon^2};
    \label{R2stu}\\
    S_3(s,t,u) &=&\frac{5s^3}{648\epsilon^3}(35s-4t-4u),\
    T_3(s,t,u) =-\frac{t^2}{432\epsilon^3}(66t-5u),\nonumber \\
    &&U_3(s,t,u) =-\frac{u^2}{432\epsilon^3}(66u-5t), 
     \label{R3stu}
\end{eqnarray}
 where $s,t$ and $u$ are the usual Mandelstam variables defined for the configuration when all external momenta are directed inside the diagram, i.e. $s=(p_1+p_2)^2$, $t=(p_2+p_3)^2$ and $u=(p_1+p_3)^2$. The total divergence is given by the sum of three channels:

\begin{equation}
   A_n(s,t,u) = S_n(s,t,u)+T_n(s,t,u)+U_n(s,t,u)\quad.
   \label{Ak}
\end{equation}

The coefficient of the highest divergence in two loops~(\ref{R2stu}) can be obtained by integrating the one-loop term over the remaining one-loop diagram, as shown in Figure~\ref{rec}, and in this case, only the first linear term works. Introducing the integral over the Feynman parameter $x$ and taking into account some complication associated with the numerators in the Feynman spinor rules, we have (omitting the obvious multiplier $1/\epsilon^2$):

\begin{equation*}
\begin{aligned}
            2S_2(s,t,u) &= 2\int_0^1 dx\,x(1-x)\left(s-t\left(1+t'\frac{d}{dt'}\right)\right)A_1(s,t',-s-t')|_{t'\rightarrow-sx}\\
            &+ 2\int_0^1 dx\,x(1-x)\left(s-u\left(1+u'\frac{d}{du'}\right)\right)A_1(s,-s-u',u')|_{u'\rightarrow-sx},
\end{aligned}
    \label{RS2}
\end{equation*}

\begin{equation*}
\begin{aligned}
            2T_2(s,t,u) = -2\int_0^1 dx\,x(1-x)\left(2t+ts'\frac{d}{ds'}\right)A_1(s',t,-s'-t)|_{s'\rightarrow-tx},
\end{aligned}
    \label{RT2}
\end{equation*}

\begin{equation*}
\begin{aligned}
            2U_2(s,t,u) = -2\int_0^1 dx\,x(1-x)\left(2u+us'\frac{d}{ds'}\right)A_1(s',-s'-u,u)|_{s'\rightarrow-ux}.
\end{aligned}
    \label{RU2}
\end{equation*}

It is easy to verify that by calculating the obtained integrals, we reproduce two-loop divergences~(\ref{R2stu}).

To calculate the three-loop coefficient~(\ref{R3stu}), one needs to add a nonlinear part to the recurrence relations by integrating two one-loop counter-terms through a one-loop diagram, as follows from the second term in Figure~\ref{rec}. In the linear part, you need to substitute the previously obtained coefficient $A_2(s,t,u)=S_2(s,t,u)+T_2(s,t,u)+U_2(s,t,u)$.

In the general case, for an $n$-loop diagram, recurrence relations for higher divergences have the following form~\cite{Borlakov}: 
\begin{equation}
\begin{aligned}
            &nS_n(s,t,u) = \\
            &=\!\int_{0}^{1}\!\!\!dx \sum_{k=0}^{n-1}\sum_{p=0}^{k}\frac{[s(-s-t)]^p[ x(1-x)]^{p+1}}{p!(p+1)!}\left(s-t\left((p+1)+t'\frac{d}{dt'}\right)\right)\times\\
            &\hspace{5.6em}\times\frac{d^p A_k(s,t',-s-t')}{dt'^p}\frac{d^p A_{n-1-k}(s,t',-s-t')}{dt'^p} \mid_{t'\rightarrow-sx}\\
            &+\!\int_{0}^{1}\!\!\!dx \sum_{k=0}^{n-1}\sum_{p=0}^{k}\frac{[s(-s-u)]^p[ x(1-x)]^{p+1}}{p!(p+1)!}\left(s-u\left((p+1)+u'\frac{d}{du'}\right)\right)\times
            \\&\hspace{5.6em}\times\frac{d^p A_k(s,-s-u',u')}{du'^p}\frac{d^p A_{n-1-k}(s,-s-u',u')}{du'^p}\mid_{u'\rightarrow-sx},
\end{aligned}
    \label{RSn}
\end{equation}
\begin{equation}
\begin{aligned}
            &nT_n(s,t,u)=\\ 
            &=-\!\int_{0}^{1}\! dx \sum_{k=0}^{n-1}\sum_{p=0}^{k}\frac{[t(-s-t)]^p[ x(1-x)]^{p+1}}{p!(p+1)!}\left(t\left((p+2)+s'\frac{d}{ds'}\right)\right)\times\\            
            &\hspace{5.6em}\times
            \frac{d^p A_k(s',t,-s'-t)}{ds'^p}\frac{d^p A_{n-1-k}(s',t,-s'-t)}{ds'^p}\mid_{s'\rightarrow-tx},
\end{aligned}
    \label{RTn}
\end{equation}
\begin{equation}
\begin{aligned}
            &nU_n(s,t,u)=\\
            &-\!\int_{0}^{1}\!\!\!dx \sum_{k=0}^{n-1}\sum_{p=0}^{k}\frac{[u(-s-u)]^p[ x(1-x)]^{p+1}}{p!(p+1)!}\left(u\left((p+2)+s'\frac{d}{ds'}\right)\right)\times\\
            &\hspace{5.6em}\times
            \frac{d^p A_k(s',-s'-u,u)}{ds'^p}\frac{d^p A_{n-1-k}(s',-s'-u,u)}{ds'^p}\mid_{s'\rightarrow-ux}.
\end{aligned}
    \label{RUn}
\end{equation}

For an amplitude proportional to the tree-level structure $\langle 14\rangle [32]$, one can write relations similar to~(\ref{RSn}-\ref{RUn}) using the above-mentioned rules for substituting momenta and changing the sign. 

With the obtained recurrence relations it is possible to calculate the leading divergences in any order of perturbation theory, starting with the one-loop divergence, without direct calculation of Feynman diagrams.

Consider now the $V-A$ operator.  In this case, the amplitude is proportional to the single spinor structure $\langle 13\rangle [42]$, and the recurrence relations have  the following form \cite{Borlakov}:
\begin{equation}
\begin{aligned}
          nS_n(s,t,u) = &-8s\!\int_{0}^{1}\!\!\! dx \sum_{k=0}^{n-1}\sum_{p=0}^{k}\frac{[s(-s-u)]^p[x(1-x)]^{p+1}}{p!(p+1)!(p+2)^{-1}}\times\\
        &\times\frac{d^p A_{k}(s,-s-u',u')}{du'^p}\frac{d^p A_{n-1-k}(s,-s-u',u')}{du'^p}\mid_{u'\rightarrow-sx},
\end{aligned}
    \label{Sn}
\end{equation}
\begin{equation}
\begin{aligned}
          nT_n(s,t,u) = &-8t\!\int_{0}^{1}\!\!\! dx \sum_{k=0}^{n-1}\sum_{p=0}^{k}\frac{[t(-t-u)]^p[x(1-x)]^{p+1}}{p!(p+1)!(p+2)^{-1}}\times\\
        &\times\frac{d^p A_{k}(-t-u',t,u')}{du'^p}\frac{d^p A_{n-1-k}(-t-u',t,u')}{du'^p}\mid_{u'\rightarrow-tx}  ,
\end{aligned}
    \label{Tn}
\end{equation}
\begin{equation}
\begin{aligned}
         nU_n(s,t,u) &= 8u\!\int_{0}^{1}\!\!\! dx \sum_{k=0}^{n-1}\sum_{p=0}^{k}\frac{[u(-s-u)]^p[x(1-x)]^{p+1}}{p!(p+1)!(p+3)^{-1}}\times\\
         &\times\frac{d^p A_{k}(s',-s'-u,u)}{ds'^p}\frac{d^p A_{n-1-k}(s',-s'-u,u)}{ds'^p}\mid_{s'\rightarrow-ux}\\ 
         &+8u\!\int_{0}^{1}\!\!\! dx \sum_{k=0}^{n-1}\sum_{p=0}^{k}\frac{[u(-t-u)]^p[x(1-x)]^{p+1}}{p!(p+1)!(p+3)^{-1}}\times\\
         &\times\frac{d^p A_{k}(-t'-u,t',u)}{dt'^p}\frac{d^p A_{n-1-k}(-t'-u,t',u)}{dt'^p}\mid_{t'\rightarrow-ux}.
\end{aligned}
    \label{Un}
\end{equation}
They allow one, as in the case of the unit operator, to find all the higher divergences based on the one-loop contribution
\begin{equation}
 S_1(s,t,u) =-\frac{8s}{3\epsilon},\ T_1(s,t,u) =-\frac{8t}{3\epsilon},\ U_1(s,t,u) =\frac{8u}{\epsilon}\quad.      
 \label{1stu}
\end{equation}

Let us now proceed to the derivation of generalized renormalization group equations which will allow one to find the high-energy asymptotics of the theories under consideration. 
For this purpose, let us construct the function

\begin{equation}
A(s,t,u)=\sum_{n=0}^\infty A_n(s,t,u)(-z)^n\quad,
\end{equation}

where $z = G/\epsilon$ and $A_0 = 1$. Multiplying equations~(\ref{RSn}-\ref{RUn}) by $(-z)^{n-1}$ and summing up from $n = 1$ to $\infty$, we obtain the corresponding differential equation.

Starting again with the case of the unit operator, one gets
\begin{equation}
    \begin{aligned}
            &-\frac{dA(s,t,u)}{dz}=
            \\
            &=\!\int_{0}^{1}\!\!\!dx \sum_{p=0}^{\infty}\frac{[s(-s\!-\!t)]^p[ x(1\!-\!x)]^{p+1}}{p!(p+1)!}(s\!-\!t((p\!+\!1)\!+\!t'\frac{d}{dt'}))(\frac{d^p A(s,t',-s\!-\!t')}{dt'^p})^2
            \\
            &+\!\int_{0}^{1}\!\!\!dx \sum_{p=0}^{\infty}\frac{[s(\!-s\!-\!u)]^p[ x(1\!-\!x)]^{p+1}}{p!(p+1)!}(s\!\!-\!\!u((p\!+\!1)\!+\!u'\frac{d}{du'}))(\frac{d^p A(s,\!-s\!-\!u',u')}{du'^p})^2
            \\
            &-\!\int_{0}^{1}\!\!\!dx \sum_{p=0}^{\infty}\frac{[t(-s-t)]^p[ x(1-x)]^{p+1}}{p!(p+1)!}(t((p\!+\!2)\!+\!s'\frac{d}{ds'}))(\frac{d^p A(s',t,-s'-t)}{ds'^p})^2
            \\
            &-\!\int_{0}^{1}\!\!\!dx \sum_{p=0}^{\infty}\frac{[u(-s-u)]^p[ x(1-x)]^{p+1}}{p!(p+1)!}u((p\!+\!2)\!+\!s'\frac{d}{ds'})(\frac{d^p A(s',-s'-u,u)}{ds'^p})^2,
    \end{aligned}
    \label{EqGr}
\end{equation}
\\
where $s', t'$ and $u'$ are the same substitutions as in eq.(\ref{RSn})-(\ref{RUn}).  The corresponding equation for the structure $\langle 14\rangle [32]$ is obtained by replacing the momenta $p_2\leftrightarrow p_4$ and changing the sign.

For the $V-A$ operator, the equation for the amplitude proportional to the structure of $\langle 13\rangle[42]$ follows from the recurrence relations~(\ref{RSn}-\ref{RUn}):

\begin{equation}
    \begin{aligned}
            &-\frac{dA(s,t,u)}{dz}=
            \\
            &=\!-8s\!\int_{0}^{1}\!\!\! dx \sum_{p=0}^{\infty}\frac{[s(-s-u)]^p[x(1-x)]^{p+1}}{p!(p+1)!(p+2)^{-1}}\left(\frac{d^p A(s,-s-u',u')}{du'^p}\right)^2
            \\
            &\hspace{1em}-8t\!\int_{0}^{1}\!\!\! dx \sum_{p=0}^{\infty}\frac{[t(-t-u)]^p[x(1-x)]^{p+1}}{p!(p+1)!(p+2)^{-1}}\left(\frac{d^p A(-t-u',t,u')}{du'^p}\right)^2
            \\
            &\hspace{1em}+\!8u\!\int_{0}^{1}\!\!\! dx \sum_{p=0}^{\infty}\frac{[u(-s-u)]^p[x(1-x)]^{p+1}}{p!(p+1)!(p+3)^{-1}}\left(\frac{d^p A(s',-s'-u,u)}{ds'^p}\right)^2
            \\
            &\hspace{1em}+\!8u\!\int_{0}^{1}\!\!\! dx \sum_{p=0}^{\infty}\frac{[u(-t-u)]^p[x(1-x)]^{p+1}}{p!(p+1)!(p+3)^{-1}}\left(\frac{d^p A(-t'-u,t',u)}{dt'^p}\right)^2.
            \\
    \end{aligned}
    \label{EqA}
\end{equation}

Here $s', t'$ and $u'$ repeat the same from eq.(\ref{Sn})-(\ref{Un}) In the next section, based on the recurrence relations~(\ref{RSn})-(\ref{RUn}) and~(\ref{Sn})-(\ref{Un}), we analyze the behaviour of the perturbation theory series and also obtain a numerical solution to the renormalization group equations~(\ref{EqGr}) and~(\ref{EqA}) in the asymptotic regime.

\section{The asymptotic behaviour of the scattering amplitude}

\subsection{The unit operator case}

As mentioned above, the recurrence relations derived in Section~3 allow one to calculate ultraviolet divergences in an arbitrary order of perturbation theory. To do this, one has to create a simple loop that allows calculating each subsequent value of the functions (\ref{Ak}) based on the previous ones.

 Let us start with relations~(\ref{Ak}) for the case of the unit operator. As a demonstration of the procedure, we present the leading divergent terms up to the fourth order:
\begin{equation}
    \begin{aligned}    
    L.P. &= 1 - \frac{z}{6} (2 s - 3 (t + u)) + 
 \frac{5z^2}{72} (4 s^2 - 2 s (t + u) + 3 (t^2 + u^2)) 
 \\&- \frac{z^3}{1296}(256 s^3 - 
    134 s^2 (t + u) + 15 s (t^2 + u^2) - 
    183 (t^3 + u^3)) 
    \\&+ \frac{z^4}{51840}(7048 s^4 - 3867 s^3 (t + u) - 
    450 s (t^3 + u^3) + 
    4789 (t^4 + u^4)) 
    \\&-\frac{z^5}{27216000}(2507480 s^5 - 1455483 s^4 (t + u) - 
   76500 s^3 (t^2 + u^2)
   \\&- 67500 s^2 (t^3 + u^3) + 
   246700 s (t^4 + u^4) - 1581108 (t^5 + u^5)) + ...\quad.
    \end{aligned}
    \label{LPu}
\end{equation}

\begin{figure}[ht]
\vspace{0.5cm}
\center{\includegraphics[scale=0.9]{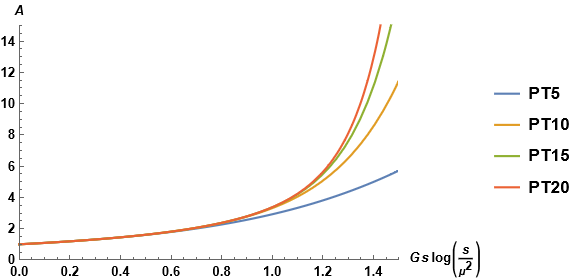}}
\caption{The dependence of the amplitude $A(y)$ on energy in different approximations of perturbation theory.}
\label{GRpt}
\end{figure}

In order to find the dependence of the scattering amplitude on energy, we use the fact that the coefficients of the leading logarithm $\log^n\mu^2$ coincide with the one of the leading poles $1/\epsilon^n$. Therefore, to find the high-energy asymptotics in the regime $s=4 E^2, t=u=-2E^2$ in the center of mass frame, it is necessary to replace $z=G/\epsilon\rightarrow -G\log(s/\mu^2)$ in eq.~(\ref{LPu}). Then the dependence of the scattering amplitude on energy in terms of the dimensionless variable $y=Gs\log s/\mu^2$ has the form shown in Figure~\ref{GRpt}, where different curves correspond  the different orders of perturbation theory. As can be seen, one has  a sign non-alternating series with increasing contributions. Thus, the amplitude increases rapidly with energy and probably has a pole at finite energy, as it is in the case of asymptotically non-free theories.

Now let us proceed to the study of equation~(\ref{EqGr}). Since an analytical solution of such complicated integro-differential equation is not possible, we  turn to numerical methods. In order to make these equations more suitable for numerical analysis, it is usefull to replace the operation of infinite summation with integration. We use the following formal technique: consider the Taylor expansion of some function $f(A+Be^{\pm i\tau})$ 
\begin{equation}
f(A+Be^{\pm i\tau})=\sum_{k=0}^\infty \frac{B^k e^{\pm i\tau k}}{k!}\frac{d^kf(A)}{dA^k}\quad.
\end{equation}
Now keeping in mind the orthogonality condition for periodic functions:
\begin{equation}
   \frac{1}{2\pi} \int_{-\pi}^\pi e^{i\tau n}e^{-i\tau m}d\tau=\delta_{mn}\quad,
\end{equation}
as well as the formula
\begin{equation}
    \int_0^1 d\xi (1-\xi)^k\frac{\xi^p}{p!} = \frac{k!}{(p+k+1)!}\quad,
\end{equation}
we obtain the following useful relation:
\begin{equation}
    \sum_{p=0}^{\infty} \frac{(BC)^pk!}{p!(p+k+1)!}\left[\frac{d^p f(A)}{dA^p}\right]^2=\frac{1}{2\pi}\int_{-\pi}^\pi \!\!\!d\tau \!\int_0^1\!\!\!d\xi(1\!-\!\xi)^kf(A\!+\!e^{i\tau}B\xi)f(A\!+\!e^{-i\tau}C)\quad.
    \label{BC}
\end{equation}
This relation replaces the infinite summation in equation~(\ref{EqGr}) with integration, which is convenient for its numerical solution.

\begin{figure}[ht]
\center{\includegraphics[scale=0.9]{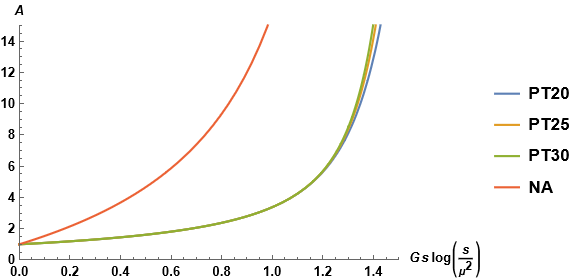}}
\caption{Comparative analysis of the function $A(y)$: NA is the numerical solution and PT is a finite number of terms of perturbation theory}
\label{GRptanal}
\end{figure}

Figure~\ref{GRptanal} shows the numerical solution of equation~(\ref{EqGr}) with the already mentioned substitution $z=G/\epsilon\rightarrow -G\log(s^2/\mu^2)$ in the same kinematics as in the case of PT series. For comparison, we also show the results of taking into account 20, 25 and 30 terms of the perturbation theory series.

As follows from the numerical solution, the scattering amplitude obtained by summing up the leading logarithms in all orders of perturbation theory increases with energy even faster than the sum of the higher order terms of perturbation theory and demonstrates a trend to the pole.

Thus, we conclude that the leading ultraviolet behaviour of Fermi theory in the case of the unit operator  is characterized by the presence of a Landau pole, and, as a result, leads to a violation of unitarity.

\subsection{The V-A operator case}

Turning to the analysis of the scattering amplitude behaviour for the V-A operator, we start by considering the leading poles of perturbation theory series that are given by the following expression~\cite{Borlakov}:
\begin{equation}
    \begin{aligned}    
    L.P. &= 1 + \frac{8z}{3} (s + t - 3 u) + 
 \frac{64z^2}{9} (s^2 + t^2 + 6 u^2)
 \\&+ \frac{256z^3}{405}(37 s^3 + 37 t^3 - 6 s^2 u - 6 t^2 u - 318 u^3) 
    \\&+ \frac{512z^4}{6075}(799 s^4 + 799 t^4 - 225 s^3 u - 225 t^3 u + 10986 u^4)
    \\&+\frac{2048z^5}{212625}(20629 s^5 + 20629 t^5 - 10064 s^4 u - 10064 t^4 u +980 s^3 u^2
   \\&+ 980 t^3 u^2 - 100 s^2 u^3 - 100 t^2 u^3 - 120 s u^4 - 120 t u^4 - 
 438184 u^5) + ...\quad.
    \end{aligned}
    \label{LPVA}
\end{equation}

\begin{figure}[ht]
\center{\includegraphics[scale=0.8]{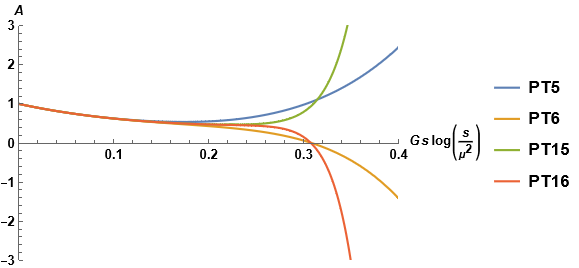}}
\caption{The behaviour of the function $A(y)$ taking into account different numbers of terms of perturbation theory.}
\label{GApt}
\end{figure}

As before, replacing $z=G/\epsilon \rightarrow -G \log(s/\mu^2)$ and moving on to the dimensionless variable $y=Gs\log (s/\mu^2)$, we  notice that the series~(\ref{LPVA}) in this case turns out to be alternating and, depending on the consideration of an even or odd number of terms, leads to either increasing or decreasing function, as shown in Figure~\ref{GApt}. Thus, based on the analysis of a finite number of terms of the series, it is impossible to make a conclusion about the asymptotic behaviour of the amplitude.

\begin{figure}[ht]
\center{\includegraphics[scale=0.6]{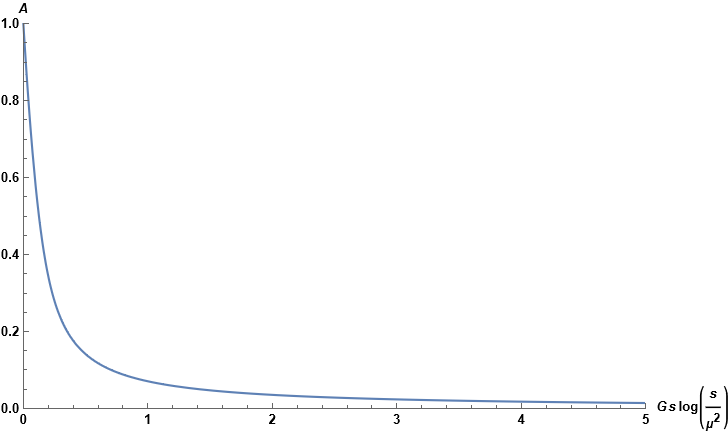}}
\caption{The function $A(y)$ as a numerical solution of equation (\ref{EqA})}
\label{GAlla}
\end{figure}

Therefore, the numerical solution of equation~(\ref{EqA}), which sums an infinite number of terms of perturbation theory, becomes important. The results of numerical solution of equation(\ref{EqA}) is shown in Figure~\ref{GAlla}.

As can be seen, the resulting solution demonstrates a rapid decrease in the amplitude at high energies. For different energy intervals, we have obtained the following numerical approximations of the function $A(y)$ with different accuracy ($h$ is the step along $y$ axis, $R^2$ is the confidence value of the approximation):
\begin{equation}
    \begin{aligned}
        &A(y) = 0.065y^{-0.91}, \hspace{0.85em} (y\in[0.001,5],\,\,\,\,h=0.001,\,R^2=0.9761),\\
        &A(y) = 0.065y^{-0.948}, \hspace{0.5em} (y\in[0.001,10],\,h=0.001,\,R^2=0.9873),\\
        &A(y) = 0.067y^{-0.987}, \hspace{0.5em} (y\in[0.001,50],\,h=0.001,\,R^2=0.9973).
    \end{aligned}
\end{equation}

Thus, we conclude that the function $A(y)$ on a large energy scale has the following asymptotic expression:
\begin{equation}
  A(y) \approx \frac{k}{y}=\frac{k}{Gs\log (s/\mu^2)},
  \label{approx}
\end{equation}
where $k\approx1/15$. The consequence of this asymptotic behaviour is analyzed in the next section.

\section{Summary}

Thus, we have demonstrated that, firstly, in a theory with a non-renormalizable interaction, it is possible not only to calculate Feynman diagrams but also to sum an infinite sequence of leading logarithms that determine the asymptotic behaviour of the scattering amplitude at high energies. This is achieved by deriving and solving the corresponding renormalization group equations, which in this case have an integro-differential form. Of course, we do not solve here the problem of fixing the infinite arbitrariness of the subtracting procedure in non-renormalizable theories, but simply assume that the divergencies are substracted in one way or another. However,  and this is essential, the  leading logarithms do not depend on this arbitrariness.

Secondly, the solution of the renormalization group equations can significantly change the asymptotic behaviour described by a finite number of terms of the perturbation theory series and differs substantially for different types of interactions, in this case, for different operators in the Lagrangian. In the case of a unit operator, our analysis indicates a rapid increase in the amplitude with energy, possibly even characterized by the presence of a pole at finite energy, leading a further violation of unitarity, which already happens in the tree-level approximation. In the case of the V-A operator, the behaviour of the amplitude changes significantly, the PT series becomes alternating, and the solution of the renormalization group equation is characterized by asymptotically free behavior!

The latter circumstance leads to a far-reaching physical consequence, which is directly related to the violation of unitarity in Fermi theory. We take a closer look at this issue.
The unitarity condition of an S-matrix is usually written as
\begin{equation}
    S^{\dagger}S=1\quad.
\end{equation}
By introducing the transition matrix $T$, which describes the interaction and using the relation~\cite{peskin}
\begin{equation}
    S=1+iT\quad,
\end{equation}
the unitarity condition takes the form
\begin{equation}
    -i(T-T^{\dagger})=T^{\dagger}T
\end{equation}
or
\begin{equation}
  Im(T)=\frac{1}{2}|T|^2\quad.
  \label{opt}
\end{equation}

This relation is usually called the optical theorem~\cite{Schwartz}. The solution of equation~(\ref{opt}) on a complex plane leads to the boundaties on the scattering amplitude:
\begin{equation}
|T|\le 2, \hspace{2em} 0\le Im(T) \le 2, \hspace{2em} \text{and} \hspace{2em} |Re(T)|\le 1\quad.
\label{ub}
\end{equation}

Therefore, in the high-energy regime, the differential
scattering cross-section in the center-of-mass frame should be limited [13]:
\begin{equation}
    \left(\frac{d\sigma}{d\Omega}\right)_{CM}=\frac{|T|^2}{64\pi^2E^2}=\frac{|T|^2}{16\pi^2s}\le\frac{1}{4\pi^2s}\quad.
\end{equation}

In the case under consideration, the differential scattering cross-section in the tree approximation is equal to:
\begin{equation}
     \frac{d\sigma}{d\Omega}=\frac{G^2s^2}{16\pi^2s}=\frac{G^2s}{16\pi^2}\quad,
\end{equation}
which leads to a violation of unitarity
for $s>2/G$.

However, as it was shown above, summation of  the leadimg  logarithms leads to an asymptotic behaviour of the amplitude that is described with fairly good accuracy by formula (\ref{approx}).
With this in mind, we have
\begin{equation}
    \frac{d\sigma}{d\Omega}\approx\frac{G^2s}{16\pi^2}\frac{k^2}{G^2s^2\,\log^2(s/\mu^2)}=\frac{C^2}{s\log^2(s/\mu^2)}<\frac{1}{4\pi^2s}\quad,  
    \label{csF}
\end{equation}
where $C=k/4\pi$ is a small constant. 

Thus, summing up the leading logarithms restores the unitarity in Fermi theory at high energy.

It is instructive to compare the expression obtained for the cross section~(\ref{csF}) with that obtained in the theory with an intermediate gauge boson. In the latter case, in the tree approximation, one has~\cite{Schwartz}:
\begin{equation}
    \frac{d\sigma}{d\Omega}=\frac{1}{16\pi^2s}\frac{g^4s^2}{[s-M^2]^2}\xrightarrow{s\gg M}\frac{g^4}{16\pi^2s}\quad,
    \label{gws}
\end{equation}
and, due to the small constant of weak interaction $g$, the unitarity condition is not violated.
In this case, the higher-order
perturbation theory corrections are taken into account using the standard renormalization group method and result in replacing the coupling constant $g^2$ with an effective "running"
coupling constant
\begin{equation}   
g_{\text{eff}}^2(s)\approx\frac{1}{c\log(s/M^2)}\quad,
\end{equation}
where the constant $c$ depends on a number of weak-interacting particles and equals 19/6 in the Standard Model. In view of this, expression~(\ref{gws}) takes the form
\begin{equation}
    \frac{d\sigma}{d\Omega}\approx\frac{g^4_{\text{eff}}}{16\pi^2s}=\frac{1}{16\pi^2c^2} \frac{1}{s\log^2(s/M^2)}\quad,
\end{equation}
and essentially has the same form as eq.~(\ref{csF}).

\section*{Acknowledgements}
The authors are grateful to D.Tolkachev for his consultation on the numerical analysis and colleagues from BLTP for useful discussions.

\bibliographystyle{hunsrt.bst}
\bibliography{bibliography}

\end{document}